%% file: ehc.tex
\documentclass[sigplan]{acmart}

\usepackage{nopageno}

\usepackage[normalem]{ulem}
\usepackage{amsmath}
\usepackage{algorithm}
\usepackage{algorithmicx}
\usepackage{algpseudocode}
\usepackage{graphicx}
\usepackage{color}
\usepackage{setspace}
\usepackage[dvipsnames]{xcolor}
\usepackage{mathtools}

\usepackage{comment}
\usepackage{fancyhdr}
\usepackage{hhline}
\usepackage{array}
\usepackage{xspace}
\usepackage{url}
\usepackage{caption}
\usepackage{subfig}
\usepackage{fancyvrb}
\usepackage{siunitx}
\usepackage{booktabs}
\usepackage{pifont}
\usepackage{flushend}
\usepackage[us,12hr]{datetime}
\usepackage{dblfloatfix}

\usepackage{multirow}

\renewcommand{\baselinestretch}{1.05}

\sloppypar

\begin{document}

\input{macros}

\keywords{Memory System, Cache Replacement, Belady's MIN\vspace{-5mm}}

\title{Making Belady-Inspired Replacement Policies More Effective Using Expected Hit Count}

\newcommand{\authspace}[0]{\hspace{18pt}}
\newcommand{\authfromaffil}[0]{\hspace{1pt}}
\newcommand{\affilspace}[0]{\hspace{15pt}}


\author{
Seyed~Armin~Vakil~Ghahani\authfromaffil{}$^\ddagger{}$
\authspace{}
Sara~Mahdizadeh~Shahri\authfromaffil{}$^\ddagger{}$
\authspace{}
Mohammad~Bakhshalipour\authfromaffil{}$^\ddagger{}$$^\S{}$
\authspace{}
Pejman~Lotfi-Kamran\authfromaffil{}$^\S{}$
\authspace{}
Hamid~Sarbazi-Azad\authfromaffil{}$^\ddagger{}$$^\S{}$
}

\affiliation{\vspace{0.4cm}
$^\ddagger$\authfromaffil{}Department of Computer Engineering, Sharif University of Technology\\$^\S$\authfromaffil{}School of Computer Science, Institute for Research in Fundamental Sciences (IPM)
\vspace{0.4cm}}

\fancyhead{}

\begin{abstract}
\input{abs.tex}

\end{abstract}

\maketitle

\thispagestyle{empty}

\input{intro.tex}

\input{motiv.tex}
\input{design.tex}
\input{method.tex}
\input{eval.tex}

\input{related.tex}

\bibliographystyle{IEEEtranS}
\bibliography{ref}

\end{document}

%% file: macros.tex
\newcommand\kasraa[1]{\noindent{\color{red} {\bf \fbox{Kasraa}} {\it#1}}}
\newcommand\TODO[1]{\noindent{\color{blue} {\bf \fbox{TODO}} {\it#1}}}
\newcommand\methodname[1]{\textsc{\mbox{#1}}}
\newcommand\appname[1]{\textsf{\mbox{#1}}}
\newcommand\componentname[1]{\textsl{\mbox{#1}}}

%% file: abs.tex
Memory-intensive workloads operate on massive amounts of data that cannot be
captured by last-level caches (LLCs) of modern processors. Consequently,
processors encounter frequent off-chip misses, and hence, lose a significant
performance potential. One way to reduce the number of off-chip misses is
through using a well-behaved replacement policy in the LLC. Existing processors
employ a variation of least recently used (LRU) policy to determine a victim
for replacement. Unfortunately, there is a large gap between what LRU offers
and that of Belady's MIN, which is the optimal replacement policy. Belady's MIN
requires selecting a victim with the longest reuse distance, and hence, is
unfeasible due to the need to know the future. Consequently, Belady-inspired
replacement polices use Belady's MIN to derive an indicator to help them choose
a victim for replacement. 

In this work, we show that the indicator that is used in the state-of-the-art
Belady-inspired replacement policy is not decisive in picking a victim in a
considerable number of cases, and hence, the policy has to rely on a standard
metric (e.g., recency or frequency) to pick a victim, which is inefficient. We
observe that there exist strong correlations among the hit counts of cache
blocks in the same region of memory when Belady's MIN is the replacement
policy. Taking advantage of this observation, we propose an expected-hit-count
indicator for the memory regions and use it to improve the victim selection
mechanism of Belady-inspired replacement policies when the main indicator is
not decisive. Our proposal offers a 5.2\% performance improvement over the
baseline LRU and outperforms Hawkeye, which is the state-of-the-art replacement
policy.

%% file: intro.tex
\section{Introduction}

The ever-increasing expansion of datasets in memory-intensive applications has resulted in massive working sets beyond what can be captured by on-chip caches of modern processors~\cite{bienia:parsec, woo:splash2, keeton:performance, regmutex, huang:hibench, ferdman_cloud, ferdman_tocs, datagoogle}. As a result, processors executing such applications encounter frequent data misses, losing significant performance potentials~\cite{cont_runahead, first_order, emc, bakhshalipour2018fast, original_runahead, kayaalp2017ric, efficient_runahead, bakhshalipour2018domino, bakhshalipour2017efficient, pejman_scale_out}. Among the data misses, which frequently happen in various levels of a modern deep cache hierarchy, Last Level Cache (LLC) misses are of more importance as for every LLC miss, the off-chip DRAM should be accessed for getting the data. Off-chip accesses significantly hurt system performance due to the limited bandwidth~\cite{hardavellas2011toward, emc, lazypim, picoserver, Huh_EDS, rogers:scaling, bakhshalipour2018fast} and long latency~\cite{ferdman_cloud, ferdman_tocs, original_runahead, efficient_runahead, VPPS, bakhshalipour2018domino, bakhshalipour2017efficient} of DRAM accesses.

While LLC misses are inevitable due to large datasets of applications, not all off-chip misses are capacity misses~\cite{bienia:parsec, woo:splash2, keeton:performance, kayaalp2017ric}. One way to reduce the number of non-capacity LLC misses is through a well-behaved replacement policy. The replacement policy decides, out of all possible candidates, which one should be evicted from the cache upon arrival of a new block of data. 

\begin{figure*}[t]
 \centering
 \includegraphics[width=0.95\textwidth]{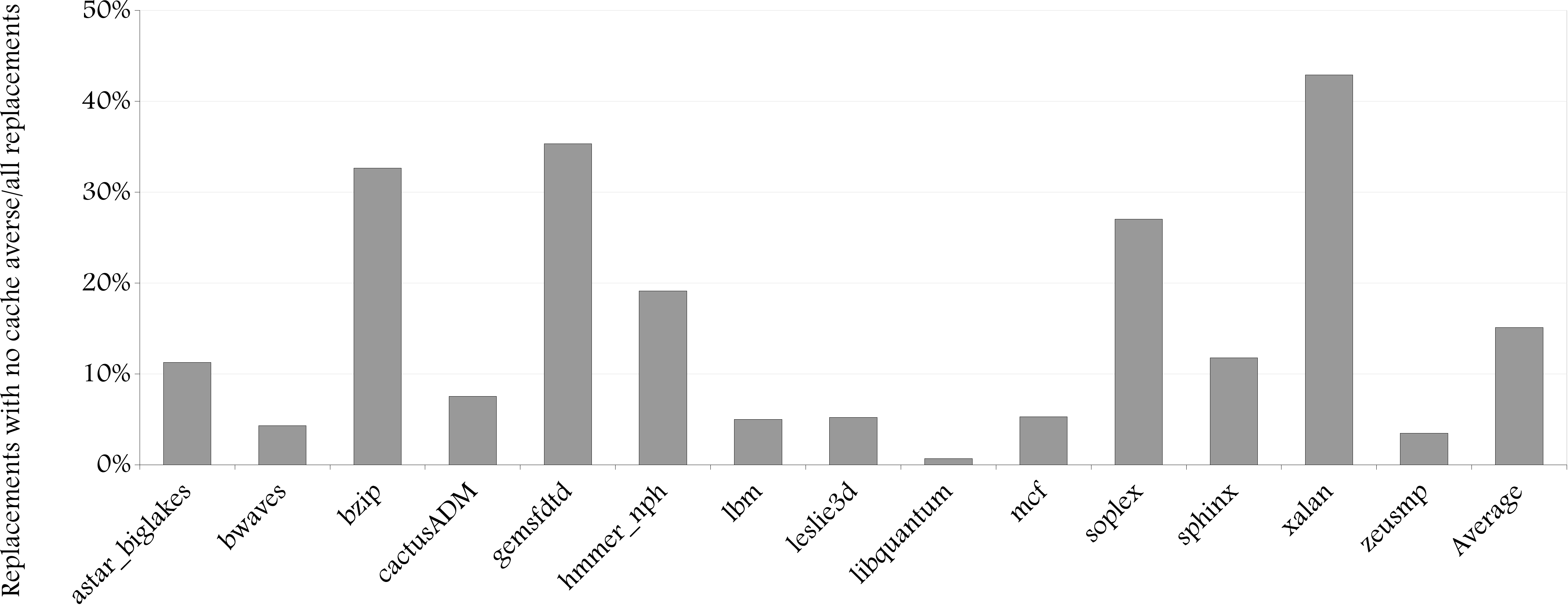}
 \caption{The fraction of replacement decisions made by Hawkeye~\cite{jain:hawkeye} when no block in the set has most-recently been touched by a cache-averse load instruction.
 \label{fig1:no_averse}}
\end{figure*}

The optimal replacement policy is \emph{Belady's MIN}~\cite{belady:replacement, mattson:replacement}. Belady's MIN replacement policy evicts a block of data that is going to be referenced further into the future. As the optimal replacement policy requires knowledge of the future, it is impractical. As a result, most replacement polices use various heuristics to evict a cache block (e.g., recency or frequency). Unfortunately, there is a significant gap between the effectiveness of Belady's MIN and practical replacement policies.

While implementation of Belady's MIN as a whole is impractical, few replacement policies emulate (approximate) Belady's algorithm to choose a victim for replacement~\cite{rajan:shepherd, jain:hawkeye}. Rajan and Ramaswamy~\cite{rajan:shepherd} used an extra storage, called a \emph{Shepherd} cache, in order \emph{not} to evict blocks until future references determine which block, based on Belady's MIN, should be chosen as the victim. Unfortunately, this technique requires a large storage to be truly effective. Jain and Lin~\cite{jain:hawkeye} benefited from the observation that with Belady's MIN, some load instructions are \emph{cache-friendly} and some others are \emph{cache-averse}. Based on this observation, they proposed \emph{Hawkeye} that uses a minimal storage to emulate Belady's MIN for the purpose of determining whether a load instruction is cache-friendly or cache-averse. Hawkeye uses cache-friendliness/-averseness of load instructions to choose a victim, i.e., evicting cache-averse blocks while maintaining cache-friendly ones. The storage requirement of Hawkeye is minimal because: (1) Hawkeye only stores the block addresses and not block data (as in Shepherd~\cite{rajan:shepherd}) to determine cache-friendliness of load instructions, and (2) Hawkeye only emulates Belady's MIN for a small number of sets in the cache, leveraging the fact that load instructions behave similarly on all cache sets.

While classifying load instructions into cache-friendly and cache-averse is quite effective\footnote{Hawkeye, which is based on classifying load instructions into cache- friendly/-averse, is the champion of the second cache replacement championship (CRC-2)~\cite{crc2:closing}.}, this technique is useful \emph{only if there are blocks that most-recently have been accessed with a cache-averse load instruction}. The blocks that are accessed with cache-averse load instructions are the prime candidate for replacement. However, \emph{if no block is most-recently accessed with a cache-averse load instruction, the replacement policy should pick a victim for replacement using standard replacement mechanisms, e.g., recency or frequency, which are in many cases ineffective}.

To show how frequently a Belady-inspired replacement policy based on cache-friendly/-averse load instructions finds no cache-averse block in a set, Figure~\ref{fig1:no_averse} shows the percentage of replacement decisions with Hawkeye when \emph{no} cache-averse block is in the set. The figure shows that for 0.5\% to 42.9\% of all replacements (with an average of 15.1\% across all benchmarks), Hawkeye should rely on traditional replacement policies to pick a victim, which limits its effectiveness. Consequently, a Belady-inspired replacement policy based on cache-friendly/-averse load instructions cannot benefit from Belady's MIN in choosing a victim in a considerable number of cases.

To address this limitation, this work makes the fundamental observation that \emph{with Belady's MIN replacement policy, there is a strong correlation among hit counts of blocks in the same memory region}\footnote{A memory region is referred to a chunk of contiguous cache blocks in the memory, holding several kilobytes of data. In this paper, we consider 128~KB memory regions.}; i.e., the hit counts of two blocks from the same memory region before being evicted by Belady's MIN are correlated. Using this observation, we estimate the hit counts of blocks in various memory regions by emulating Belady's MIN and \emph{use the expected hit count for replacement when no cache-averse block is available in a set}. Just like Hawkeye, to estimate the expected hit count of a region, we only need to store the block addresses and not block data. Moreover, as we only need to estimate the hit count of memory regions and not individual blocks, the storage overhead of our proposal is insignificant.

In this paper, we make the following contributions:

\begin{itemize}
 \item We show that a Belady-inspired replacement policy that classifies load instructions into cache-friendly or cache-averse, in a considerable fraction of cases, are unable to perform the replacement decisions based on the history of loads.
 
 \item We show that with Belady's MIN replacement policy, there is a strong correlation between the hit count of a cache block in two consecutive residencies in the cache.

 \item Furthermore, we show that with Belady's MIN replacement policy, not only there is a strong correlation between the hit count of a cache block in two consecutive residencies but also there is a strong correlation among hit counts of blocks in the same memory region. 
 
 \item Using these observations, we augment a Belady-inspired replacement policy based on cache-friendly-averse load instructions with a small structure to track the hit counts of various memory regions and use the region hit count in replacement decisions when no cache-averse block exists in a set to improve the victim selection quality.

 \item We use a simulation infrastructure to evaluate our proposal in the context of both single and multi-core processors. Our results show that our proposal offers 17.5\% lower Miss Per Kilo Instructions (MPKI) and 5.2\% higher performance, as compared to the baseline LRU, and outperforms all prior state-of-the-art replacement policies. 

\end{itemize}

%% file: motiv.tex
\section{Motivation}

In this section, we show that there is a strong correlation between the hit counts of cache blocks in the same region of memory when the replacement policy is Belady's MIN. We first show that there is a strong correlation among the hit counts of a cache block in its recent residencies in the cache when the replacement policy is Belady's MIN. We then extend the results to blocks in the same region of memory.

\begin{figure*}[b]
 \centering
 \includegraphics[width=0.95\textwidth]{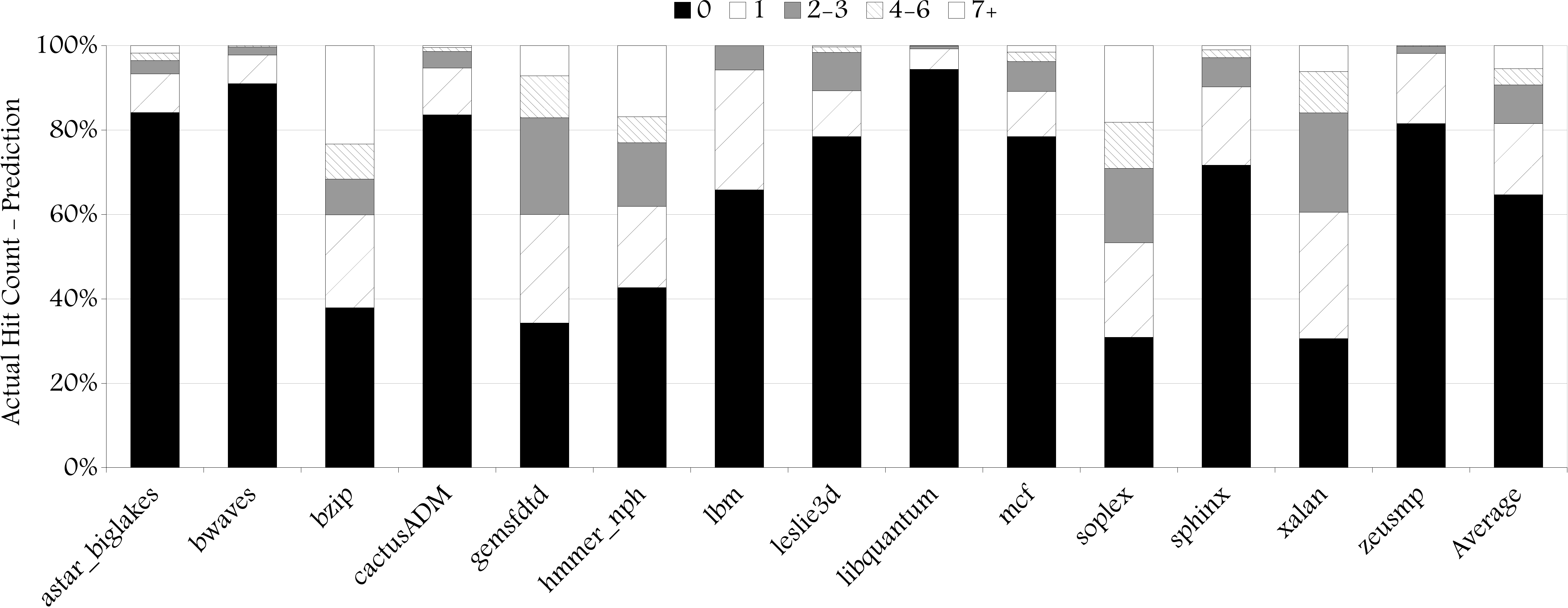}
 \caption{The difference between the actual and predicted hit count of cache blocks when the replacement policy is Belady's MIN. The predicted hit count of a cache block is the average hit count of the block in its last four residencies in the LLC. 
 \label{fig2:hc_block}}
\end{figure*}

\begin{figure*}[b]
 \centering
 \includegraphics[width=0.95\textwidth]{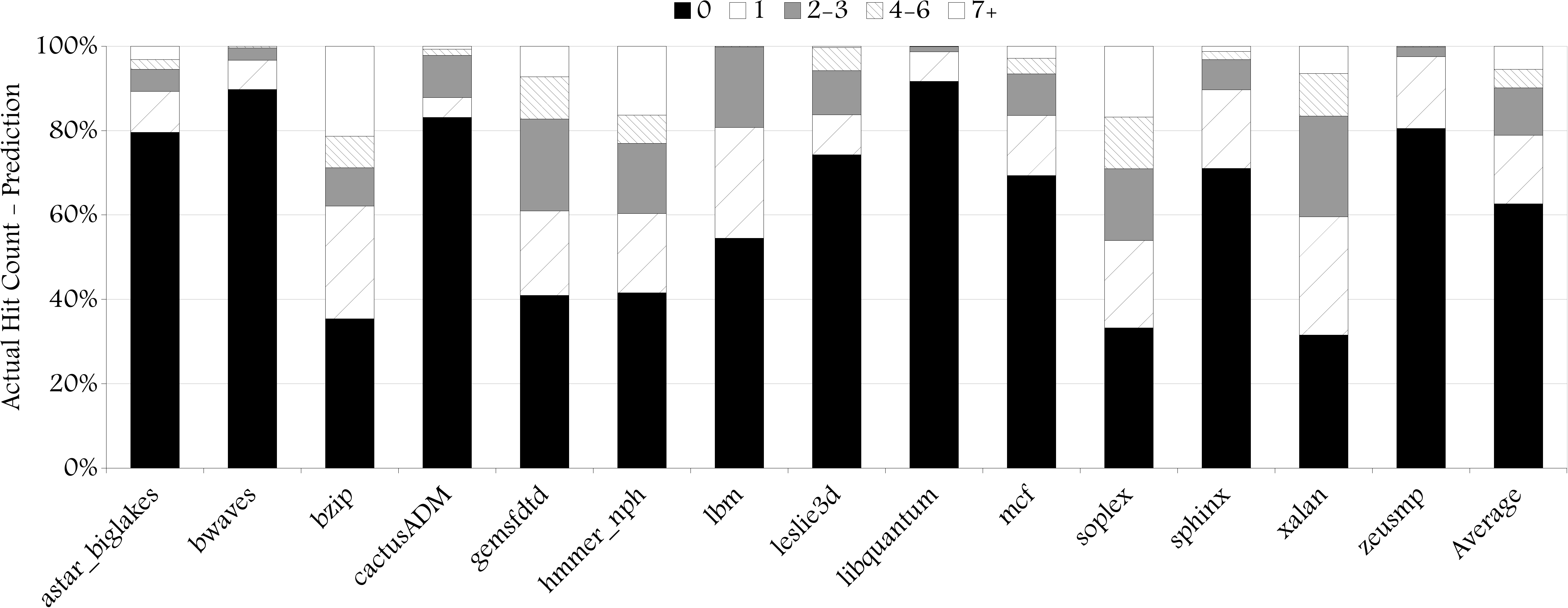}
 \caption{The difference between the actual and predicted hit count of memory regions when the replacement policy is Belady's MIN. The predicted hit count of a memory region is the average hit count of the last four recently-evicted cache blocks in the region.
 \label{fig3:hc_region}}
\end{figure*}

To measure the correlation among the hit counts of a cache block in its various residencies in the cache, we predict the number of hits of a cache block by averaging the number of hits that the block experienced in its last four residencies in the LLC. Figure~\ref{fig2:hc_block} shows the difference between the actual hit count and the prediction for several benchmarks. The details of the methodology can be found in Section~\ref{sec:method}.

As Figure~\ref{fig2:hc_block} shows, 30.6\% to 94.4\% of predictions are 100\% accurate in various benchmarks. On average across all benchmarks, 64.6\% of hit-count predictions are exact and 81.5\% of them differ with the actual hit count by at most one. Most of the other predictions are also close to the actual hit count. The figure clearly shows that there is a strong correlation among the hit counts of a cache block in its recent residencies in the LLC. The correlation is so strong that we can easily predict the hit count of a block with a high accuracy.
Not only there is a strong correlation among the hit counts of a cache block in its recent residencies in the cache, but also there is a strong correlation among the hit counts of memory regions when the replacement policy is Belady's MIN. To measure the correlation among the hit counts of memory regions, we predict the number of hits of a cache block in each region by averaging the number of hits that the last four evicted blocks in that region have experienced. Figure~\ref{fig3:hc_region} shows the difference between the actual hit count and the prediction for several benchmarks.

As shown, 31.5\% to 91.6\% of predictions are exactly correct in various benchmarks. On average, 62.6\% of predictions are correct and 78.9\% of them differ with the actual hit count by at most one. Comparing Figure~\ref{fig2:hc_block} and Figure~\ref{fig3:hc_region} reveals that prediction accuracy of block hit count using past hit counts of the cache block itself and using hit counts of blocks in the same region is very close. 

While these results are strong, in the sprite, they are an extension of the findings of prior work~\cite{wu:ship}. Wu et al.~\cite{wu:ship} showed that the re-reference interval of cache blocks is correlated with some \emph{signatures} like program counter, memory region, and instruction sequence. These signatures can be used to predict the re-reference interval of a cache block. Corroborating prior work, we observe that there is a correlation between the memory region and re-reference interval of cache blocks. Moreover, by extending the findings of prior work, we show that \emph{there is a strong correlation between memory regions and the hit counts of cache blocks}. Note that prior work concluded that the correlation between memory regions and re-reference interval is \emph{not} strong in a \emph{traditional} replacement policy. This work, however, shows that there is a strong correlation between the memory regions and the hit counts of cache blocks \emph{with Belady's MIN}. Taking advantage of this strong correlation, we attempt to improve the effectiveness of Belady-inspired replacement policies. 

Many pieces of prior work~\cite{lai:selective, lai:dead, khan:using, liu:bursts, jaleel:adaptive, jaleel:rrip, qureshi:insertion, wu:ship, vakil2018cache} enhanced replacement decisions by identifying and evicting/bypassing dead blocks. A dead block is a cache block with no further reference during its current residency in the cache (i.e., blocks with zero remaining hits). Our proposal is a generalization of prior work: instead of just relying on dead blocks (i.e., blocks with zero remaining hits), our proposal estimates the expected number of hits of a cache block (zero or larger) and benefits from that information in the decision making process. We find that prior proposals that employ a dead block predictor for making replacement decisions fundamentally suffer from one or two of the following problems: (1) whenever a live block \emph{mistakenly} identifies as dead, a cache miss is inevitable, and (2) whenever \emph{all} blocks in a set are predicted as live blocks, the replacement policy is unable to effectively choose the best victim for replacement.

%% file: design.tex
\section{Our Proposal}

Taking advantage of the strong correlation of hit counts of blocks in the same memory region, we plan to improve the effectiveness of Belady-inspired replacement policies. To make what we add to a Belady-inspired replacement policy understandable, we first briefly introduce the main components of a Belady-inspired replacement policy and then explain what we add to it.

Replacement decisions with Belady's MIN require knowledge of future accesses, but to determine whether the current access is a miss or a hit, having past accesses suffices. If a cache access turns into a miss, however, Belady's MIN requires knowledge of future accesses to replace a cache block. There are two strategies for implementing Belady-inspired replacements: (1) directly using Belady for replacement decisions, and (2) using hit/miss behavior with Belady's MIN to derive an indicator for replacement decisions. Both strategies require extra storage in addition to the cache: former strategies require knowledge of future accesses to pick a replacement candidate, and hence, need to keep all the blocks in a storage and defer the replacement decision to the future, while the latter strategies require knowledge of past cache accesses. Prior work~\cite{jain:hawkeye} showed that to get reasonable results, a Belady-inspired replacement policy requires an extra storage with enough capacity for eight times the number of blocks in the cache.

Due to the massive size of the required extra storage, techniques that directly emulate Belady's MIN (e.g., Shepherd Cache~\cite{rajan:shepherd}) are not considered efficient. Aiming to build a storage-efficient replacement policy, Jain and Lin~\cite{jain:hawkeye} work out two strategies: (1) they do not directly use Belady's MIN to derive replacement decisions; instead they derive indicators to be used in picking a victim for replacement, as this enables them not to store block data, which is much larger than the block address, and (2) they use sampling by relying on indicators that can be trained using only a small number of cache sets. 

In this work, we take the advantage of Belady's MIN to derive an indicator, i.e., the \emph{expected hit counts of memory regions}, and as such, do not need to store the block data in the extra storage. Moreover, as memory regions are much larger than cache blocks and span over a large number of cache sets, the expected hit counts of memory regions can be estimated using only a small number of cache sets. Doing so enables using sampling while estimating the expected hit counts of memory regions.

In this part, we review how a Belady-inspired replacement policy decides if an access is a hit or a miss using Belady's MIN, and then show how we can measure and use the expected hit counts of memory regions. We first explain the algorithm for determining the hit/miss decision and then describe the hardware implementation. We assume that past sequence of accesses to every set is recorded. As the goal is to determine the hit/miss events, the recorded sequence only includes block addresses and not block data. Upon a new access to the set, we attempt to find the corresponding address in the recorded sequence of addresses. If the address is not found, this is the first access to the address, and hence, is a miss. Otherwise, we need to examine the sequence of accesses between the last and current access to this address. We refer to the interval between two accesses to the same address as the \emph{reuse interval}. We need to find all the reuse intervals of other accesses that fall within the reuse interval of the current access. We then need to find the maximum number of overlaps among such intervals, which we refer to it as the \emph{occupancy}. If the occupancy reaches or exceeds the size of the cache set (i.e., cache associativity), the current access is a miss, and otherwise, a hit under Belady's MIN.

To implement this algorithm in hardware, prior work~\cite{jain:hawkeye} suggested recording the occupancy of a set after every access to the set. This means that we record both the address of the access and the occupancy of the set after the access, as part of the recorded sequence of accesses. For every new access to the cache, the recorded sequence is searched to find an access to the same address in the past. In case a match is found, if the recorded occupancy of all the recorded accesses after the match is less than the associativity of the cache, this access is a hit, and otherwise, it is a miss. In case the access is a hit, the occupancy number of all the accesses after the match is incremented to indicate that the currently accessed address was in the cache since last access. 

As there are many repetitions in the sequence of past addresses, to reduce the storage overhead, the occupancy numbers and addresses are stored in two different structures. For every set, there is an \emph{occupancy vector} that stores the sequence of occupancy numbers and an \emph{address cache} for storing the corresponding addresses. In addition to the address, each entry in the address cache also has a pointer that points to the last occurrence of this address in the occupancy vector. Depending on what indicator the Belady-inspired replacement policy trains for replacement decisions, there might be other elements in the address cache (e.g., PC of the load instruction).

In general, Belady may require recording all the past accesses to decide if the current access is a miss or a hit. However, in practice, as shown by prior work~\cite{jain:hawkeye}, for each set, recording eight times the associativity of the cache is enough to decide if an access is a hit or a miss under Belady's MIN in most of cases. Moreover, to reduce the storage requirement, the aforementioned mechanism is only performed on a small number of cache sets (e.g., one for every sixty-four sets)~\cite{qureshi:insertion}.

\subsection{Design Overview}

Our proposal, named \emph{Expected Hit Count (EHC)}, is built on top of a Belady-inspired replacement policy and attempts to exploit the expected-hit-count phenomenon. EHC requires extending the tag storage of each block in the LLC with only three bits. The storage unit, named \emph{Expected Further Hits}, is a count-down counter that shows \emph{the number of further hits that the block is expected to have in its current residency in the cache}.

\subsection{Selecting Victim}

When the trained indicator of the baseline Belady-inspired replacement policy is not decisive (e.g., all blocks in a set are cache-friendly in Hawkeye), the replacement policy cannot use load history for victim selection and instead inevitably relies on a standard policy. We use the expected hit count to improve the quality of victims in such cases. Many recent replacement policies are based on \emph{Re-Reference Interval Prediction (RRIP)}~\cite{jaleel:rrip} (e.g.,~\cite{wu:ship, jain:hawkeye, Jimenez:multi}), so in this part, we explain how to use the expected hit count to improve the victim selection mechanism of RRIP. We emphasize that EHC is not limited to RRIP and can be used with other replacement policies as well.

RRIP associates a number with each cache block named \emph{Re-Reference Prediction Value (RRPV)} and replaces a block with the highest RRPV. EHC updates the victim selection mechanism based on ``how many further hits do we expect to get from each cache block?'' Every time that a new block is brought into a set, its expected hit count is placed into \emph{Expected Further Hits} counter in the cache. The value of the expected hit count is empirically chosen to be one. For every access to this block, the content of \emph{Expected Further Hits} is decremented. When a victim needs to be selected, we combine \emph{Expected Further Hits}, which determines how many further hits we expect to see for this block, \emph{and} RRPV to choose a victim. We examine the value of `$ExpectedFurtherHits - RRPV$' for all candidates and evict the block with the lowest value\footnote{If more than one block has the lowest value, we pick the first one.}.

%% file: method.tex
\section{Methodology}
\label{sec:method}

\subsection{Simulation Infrastructure}
We evaluate our proposal using the simulation framework released by the Second Cache Replacement Championship (CRC-2)~\cite{crc2}. Table~\ref{table:method} summarizes the key elements of our methodology. We target both single-core and four-core processors with a 2~MB per-core shared LLC. The processors benefit from a non-inclusive cache hierarchy that employs LRU as the default replacement policy. We report both cache statistics and the end-to-end performance of the competing policies.

\begin{table}[t!]
\small
 \begin{center}
  \caption{Evaluation Parameters.}
  \label{table:method}
    \begin{tabular}{| l || l |}
     \hline
       \textbf{Parameter}               & \textbf{Value} \\
     \hline
     \hline
       {Processing Nodes}               & 6-stage pipeline, 256-entry ROB \\
     \hline
       \mbox{L1-D/I} Caches             & 32~KB, 8-way, 4-cycle load-to-use \\
     \hline
       Private L2 Cache                 & 256~KB, 8-way, 8-cycle access latency \\
     \hline
      Shared LLC                        & 2~MB per core, 16-way, 20-cycle hit latency \\
     \hline
     \hline
       \multirow{2}{*}{Data Prefetcher} & L1 next-line prefetcher \\
                                        & L2 PC-based stride prefetcher~\cite{stride_prefetcher} \\
       \hline
    \end{tabular}
 \end{center}
\end{table}

We use SPEC CPU2006~\cite{spec} benchmarks for evaluating the competing replacement policies. For the multi-program workloads, we randomly choose a hundred combinations of single-core programs and use them to evaluate the competing policies on a four-core processor. For single-core evaluations, we execute 4-billion instructions and use half of the instructions for warm-up and the rest for actual measurements. For the multi-core evaluations, we execute 2-billion instructions \emph{per core} and use the first half for warm-up and the rest for measurements.

\subsection{Evaluated Methods} We evaluate the following replacement policies:\\

\noindent\textbf{Baseline LRU.} Well-known \emph{Least Recently Used} replacement policy is used as the baseline in our evaluations. It keeps four bits per block in each set to establish the LRU stack. \\

\noindent\textbf{Dynamic RRIP (DRRIP)}~\cite{jaleel:rrip}. Each block has a 3-bit storage unit named RRPV. Upon each hit, RRPV of the block is set to zero, and upon each miss, the block with the maximum RRPV is evicted. If none of the blocks have the maximum RRPV, the RRPV of all blocks is incremented. This procedure is repeated until at least one block gets the maximum RRPV. It uses set-dueling for choosing an insertion policy between \emph{Static RRIP (SRRIP)} and \emph{Bimodal RRIP (BRRIP)}. In SRRIP, all blocks are inserted with RRPV of \emph{maximum minus one}. In BRRIP, the RRPV of inserted block gets the value of \emph{maximum minus one} (with the probability of $\frac{1}{32}$) or \emph{maximum} (with the probability of $\frac{31}{32}$). Thirty-two random sets emulate SRRIP and another thirty-two random sets emulate BRRIP. The remaining sets follow the winner of the duel. The total area overhead of DRRIP is 12~KB/48~KB in single-core/four-core substrate.\\

\noindent\textbf{SHiP}~\cite{wu:ship}. \emph{Signature-Based Hit Predictor} replacement policy relies on RRIP but attempts to distinguish the dead blocks from the live ones, which will be re-referenced in the cache. It uses the PC of corresponding instruction for classification of dead and live blocks. The replacement policy differs from RRIP since it sets the RRPV of a block during insertion based on the dead block prediction. SHiP uses an LRU replacement simulator to find the cache-friendliness/-averseness of blocks, and stores the outcome in a particular structure. In addition, SHiP uses two bits per block for storing RRPV information. The total area overhead of SHiP is 39~KB/156~KB in single-core/four-core substrate.\\

\begin{figure*}[b]
 \centering
 \includegraphics[width=0.9\textwidth]{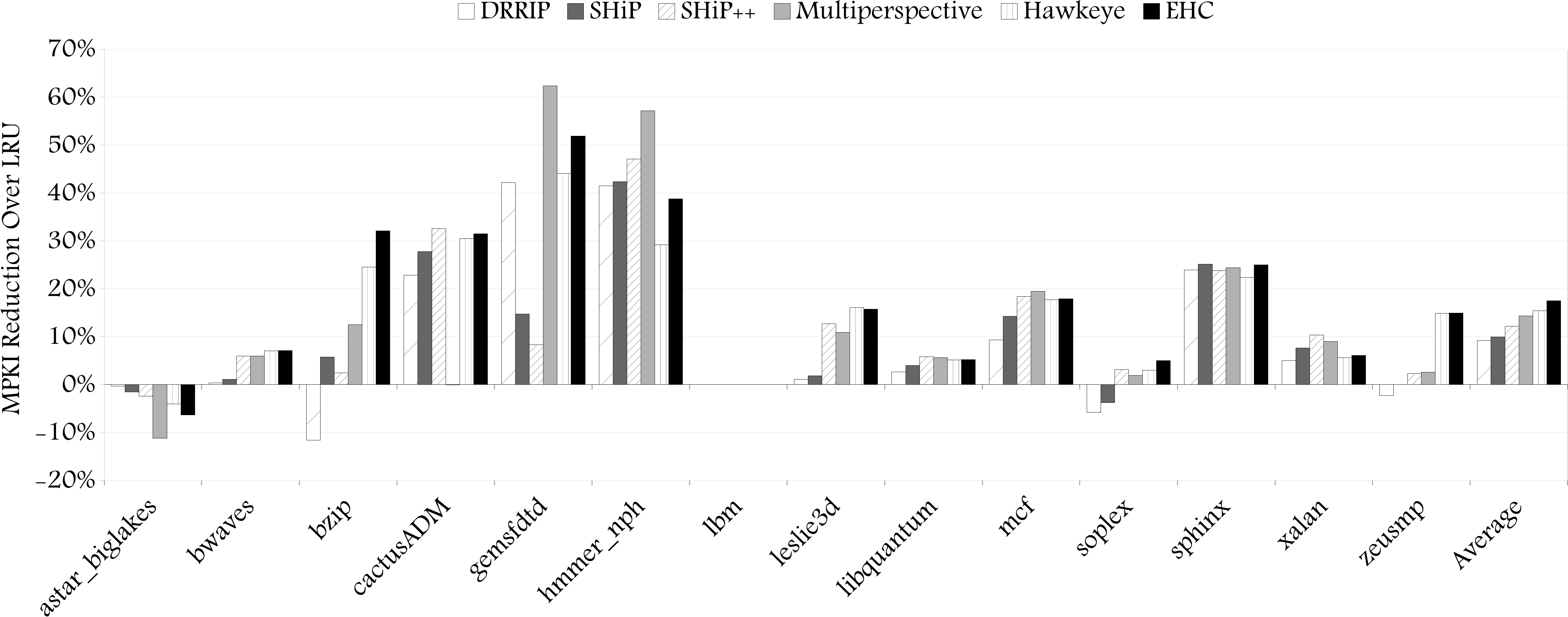}
 \caption{MPKI reduction of the competing replacement policies over the baseline LRU.
 \label{fig10:mpki}}
\end{figure*}

\noindent\textbf{Multiperspective.} This replacement policy leverages machine learning concepts to determine whether the incoming block should bypass the cache or not. It also uses a mechanism for sending feedback to assist in choosing a victim. The machine learning mechanism used in this method exploits several \emph{features} to determine the output and has a sampler unit to calculate the weight of each feature. The sampler and other metadata structures used in this mechanism occupy approximately 25.25~KB/95.5~KB storage in a single-/four-core processor. The baseline replacement of this algorithm is different in single-core and four-core systems: in single-core systems, it uses \emph{PseudoLRU}~\cite{PseudoLRU} as the baseline replacement policy of the main cache, which has 3.75~KB hardware overhead; however, in four-core systems, SRRIP~\cite{jaleel:rrip} is the baseline replacement policy, with 32~KB additional storage overhead. The total storage overhead of this method is 29~KB/127.5~KB in single-core/four-core substrate.\\

\noindent\textbf{Hawkeye}~\cite{jain:hawkeye}. Hawkeye uses an optimal replacement simulator, named \emph{OPTGen}, to simulate Belady's MIN and classify load instructions (PCs) into cache-friendly or cache-averse. OPTGen's hardware overhead is 15.2~KB. Moreover, Hawkeye uses three bits for the RRPV, which are set, upon each access based on the averseness or friendliness of the PCs of incoming accesses. Upon each miss, Hawkeye evicts a cache-averse block from the cache. If there is no cache-averse block in the set, it chooses the \emph{oldest} block in the set as the victim. The total overhead of this replacement policy is 30~KB/90~KB in single-core/four-core substrate.\\

\noindent\textbf{EHC.} Implemented on top of Hawkeye~\cite{jain:hawkeye}. This policy makes changes in the victim selection mechanism of Hawkeye to efficiently select a victim when all cache blocks in a given set are predicted to be cache-friendly. The total storage overhead of EHC on top of Hawkeye is 12~KB per core.\\

%% file: eval.tex
\begin{figure*}[]
 \centering
 \includegraphics[width=0.9\textwidth]{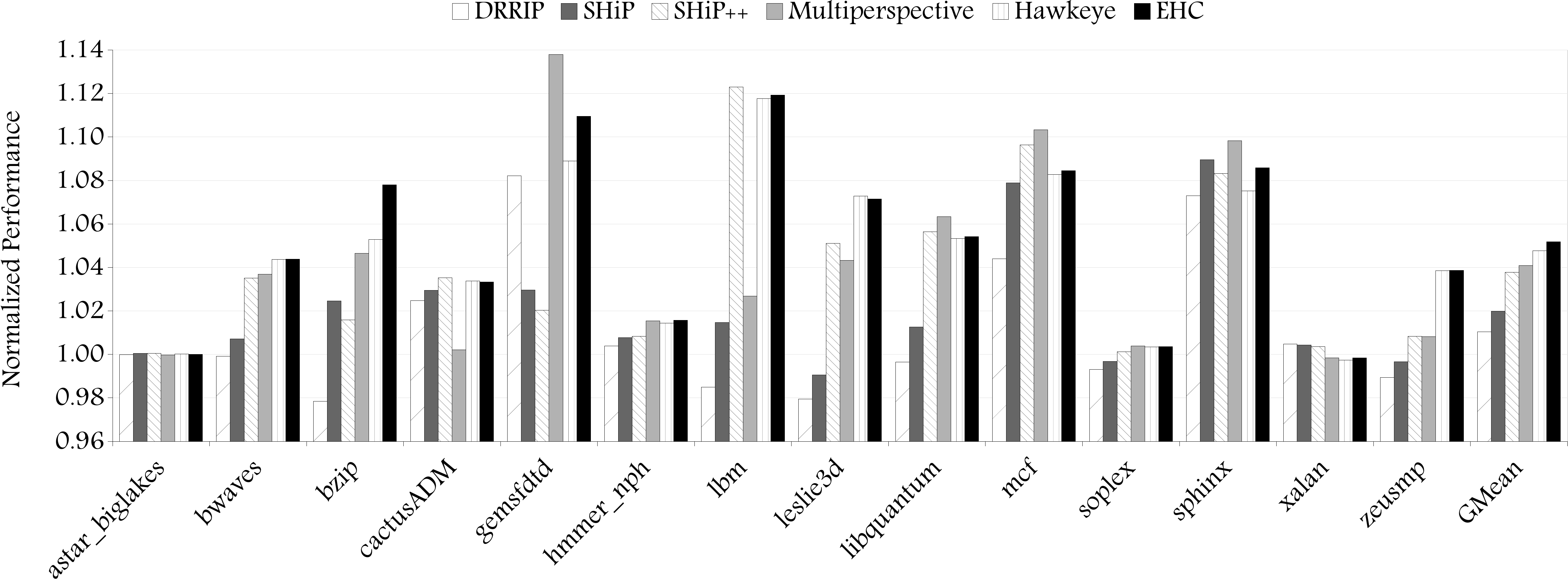}
 \caption{Performance improvement of the competing replacement policies over the baseline LRU.
 \label{fig11:single_perf}}
\end{figure*}

\section{Evaluation}
\label{sec:eval}

\subsection{Miss Reduction}

Figure~\ref{fig10:mpki} shows the MPKI reduction of various policies over the baseline LRU. As shown clearly, our proposal outperforms all previously-proposed replacement policies. On average, our proposal reduces the MPKI by 17.5\%. The second best policy is Hawkeye, which offers a 15.4\% MPKI reduction.

\subsection{Performance}

Figure~\ref{fig11:single_perf} compares the performance improvement of the evaluated policies over the baseline LRU. We use the Instruction per Cycle (IPC) as the metric for performance. EHC offers the highest performance improvement on average. The average performance improvement is 5.19\% across all workloads. The second best policy is Hawkeye with an average performance improvement of 4.76\%.

\begin{figure*}[]
 \centering
 \includegraphics[width=0.9\textwidth]{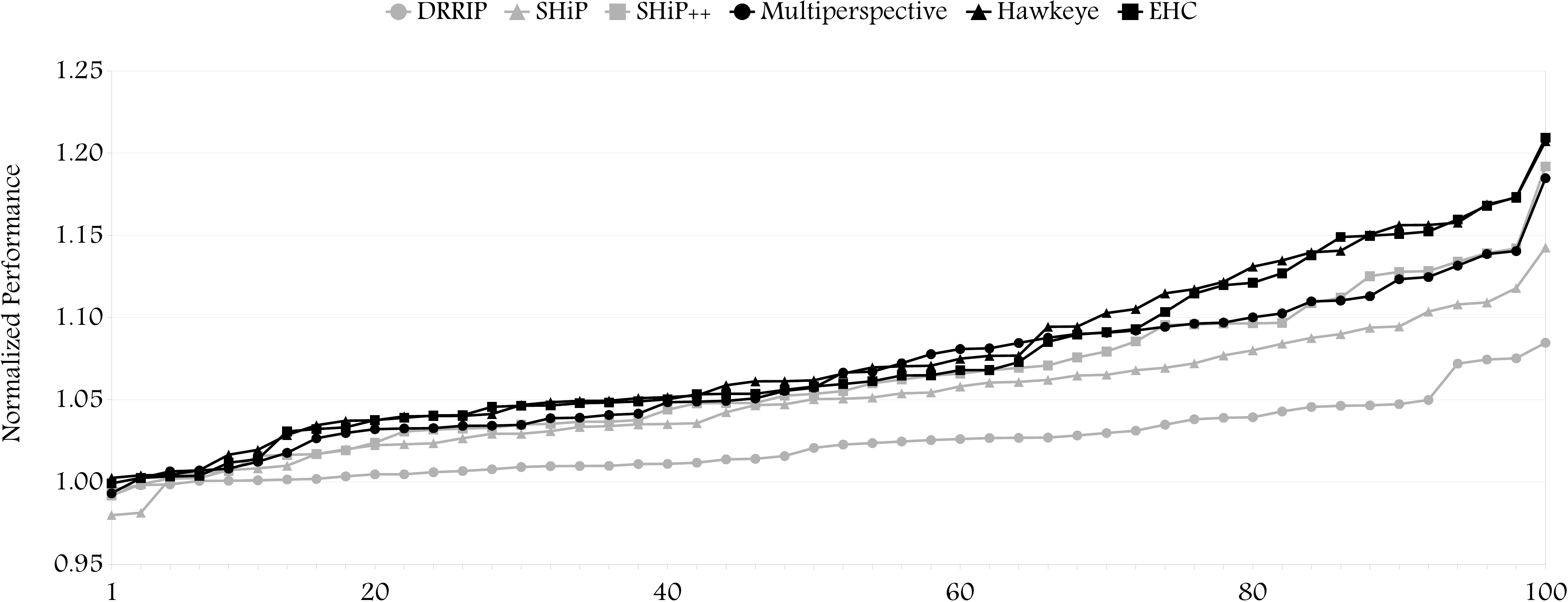}
 \caption{Performance improvement of the competing replacement policies over the baseline LRU.
 \label{fig12:multi_perf}}
\end{figure*}

Figure~\ref{fig12:multi_perf} compares the performance improvement of the evaluated policies over the baseline LRU in a four-core system. Again, EHC achieves the highest performance, outperforming all other replacement policies on average.  

\subsection{Why is EHC Effective?}
\begin{figure*}[]
 \centering
 \includegraphics[width=0.90\textwidth]{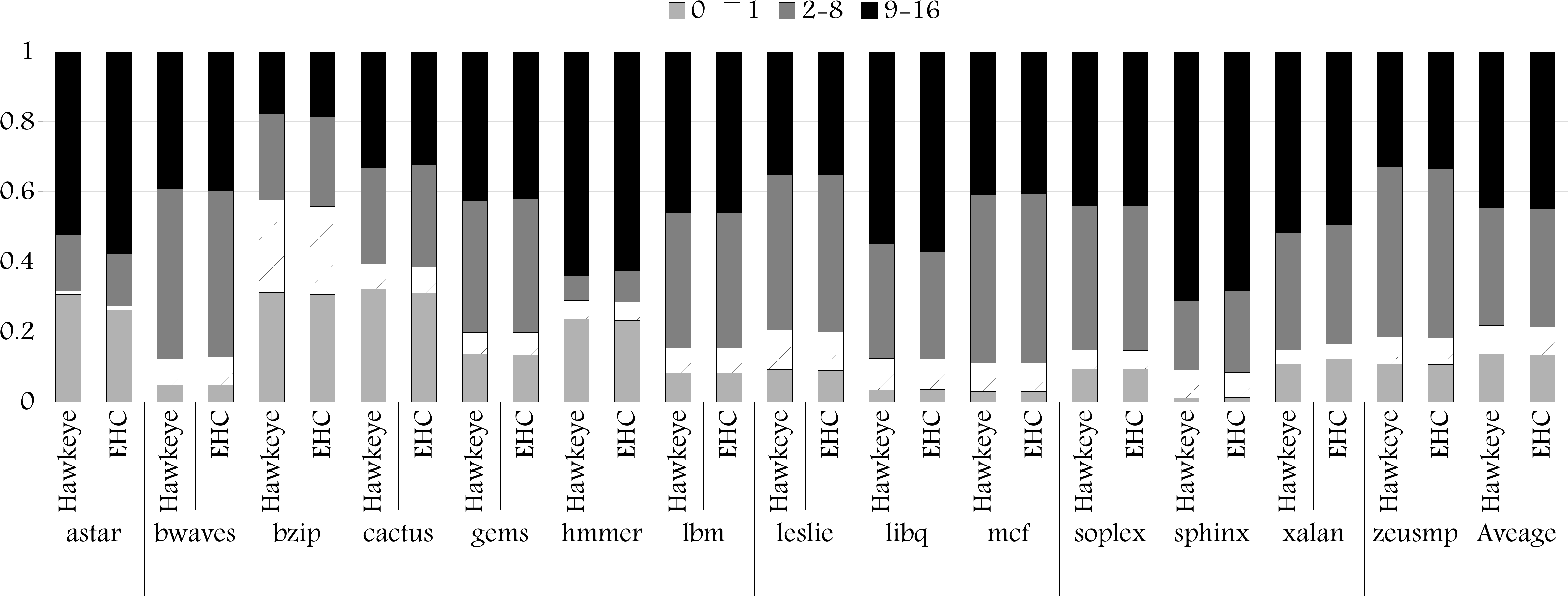}
 \caption{The quality of chosen victims in Hawkeye and EHC. Blocks in a set are sorted with 
	 the reuse distance: \textit{Block 0} has the longest and \textit{Block 16} has the
	 shortest reuse distance. Selecting a victim with the lower index is better.
 \label{fig9:belady_comparison}}
\end{figure*}

To show that the expected hit count with a non-ideal replacement policy is correlated with the reciprocal of the reuse distance, Figure~\ref{fig9:belady_comparison} compares the quality of the chosen replacement victims in Hawkeye and EHC. Every time a victim needs to be selected, we sort the blocks in the set and the incoming block based on their reuse distance. \emph{Block 0} has the longest and \emph{Block 16} has the shortest reuse distance. Ideally, a replacement policy always picks \emph{Block 0}. Comparing EHC and Hawkeye reveals that the victims chosen by EHC have higher quality as compared to those chosen by Hawkeye.

%% file: related.tex
\section{Related Works}

The replacement policy is basically a prediction mechanism to determine the best candidate among the residing blocks in a set of the cache to be replaced with the incoming block. Different replacement policies have different approaches to perform this prediction, which can be classified as follows:

\subsection{Static Prediction}
This class of replacement policies predict the best victim candidate without concerning the blocks' unique features and solely based on events such as hit, miss or eviction. Therefore, all victim selection decisions are only influenced by such short-term events, which cannot distinguish different access patterns. LRU and Pseudo LRU~\cite{PseudoLRU}, for instance, perform well in case of recency-friendly access patterns. However, they perform poorly on other types of access patterns as the assumption that the block which is referenced further in the past should be evicted could lead to cache pollution and cache thrashing in some applications. RRIP~\cite{jaleel:rrip} is effective for streaming accesses but can degrade the performance of applications that follow other patterns. Adapting these static approaches to the unique requirements of an application is useful. Techniques like Set Dueling~\cite{qureshi:insertion}, which is used in policies like DIP~\cite{qureshi:insertion} and DRRIP~\cite{jaleel:rrip}, can improve the performance of static approaches by selecting the best-fitting approach to choose the victim. TADIP~\cite{jaleel:adaptive} extends static prediction concept to shared caches. Unfortunately, adapting the replacement policy based on the workload would not fill the large gap between the performance of this class of policies and the optimal algorithm, as the winning policy would apply to all blocks regardless of the differing behavior.

\subsection{Dynamic Prediction}
The dynamic prediction is based on the unique characteristics of a block such as tag, PC, live distance or age. These replacement policies attempt to identify, categorize, and rank blocks as they enter the cache. As a result, in case of a conflict, the block that is predicted to be the one which is least likely to be re-referenced is chosen as the victim. Learning the behavior of each category, however, can be based on different information and through different mechanisms. Evicted-Address Filter (EAF)~\cite{EAF}, for example, keeps track of recently evicted blocks using a Bloom Filter. Some replacement policies leverage a Dead Block Predictor (DBP), which is a mechanism to predict the time when a block becomes dead, to evict dead blocks from the cache. Policies such as cache-burst~\cite{liu:bursts}, SDBP~\cite{SDBP}, SHiP~\cite{wu:ship}, and Leeway~\cite{Leeway} are based on DBPs and use various indicators to decide whether a block is dead. PRP~\cite{tor:prp} uses a probabilistic approach to determine the approximate probability of occurring a hit for a block according to its age. EVA~\cite{beckmann:eva} similarly uses theoretical analysis to calculate the Economic Value Added (EVA) for each block based on the block's age to determine if it is worth to keep the block in the cache. 

Multiperspective~\cite{Jimenez:multi} categorizes blocks not solely based on one or two possible features, on the contrary, it classifies blocks based on different characteristics that may have an effect on the prediction of reuse behavior of blocks. Policies such as RDP~\cite{keramidas:rdp}, Shepherd Cache~\cite{rajan:shepherd}, and Hawkeye~\cite{jain:hawkeye} emulate the Belady's optimal policy to reduce the gap between practical replacement policies and Belady's MIN. RDP~\cite{keramidas:rdp} uses an expensive content-addressable memory for storing the data of each PC. Shepherd Cache~\cite{rajan:shepherd} requires storing the block data besides the block addresses to emulate Belady's MIN, and hence, cannot effectively use its storage to simulate the Belady's algorithm. Hawkeye uses Belady's MIN to derive an indicator to help to choose victims. Consequently, it does not need to store the block data, and hence, it benefits from a wider window to keep a record of past references to emulate Belady. We showed that Hawkeye's indicator for choosing a victim is not decisive in a considerable number of cases. To address this limitation, we proposed a different indicator, the expected hit counts of memory regions, to help to choose the victim when the main indicator is not decisive. \vspace{5mm}